\documentclass{ceurart}

\usepackage{graphics}

\begin{document}

\copyrightyear{2023}
\copyrightclause{Copyright by the paper’s authors. Copying permitted only for private and academic purposes. In: M. Leyer, Wichmann, J. 
(Eds.): Proceedings of the LWDA 2023 Workshops: BIA, DB, IR, KDML and WM. Marburg, Germany, 09.-11. October 2023, published at http://ceur‐ws.org}

\conference{LWDA'23: Lernen, Wissen, Daten, Analysen.
  October 09--11, 2023, Margburg, Germany}

\title{Preliminary Results of a Scientometric Analysis of the German Information Retrieval Community 2020-2023}

\author[1]{Philipp Schaer}[%
orcid=0000-0002-8817-4632,
email=philipp.schaer@th-koeln.de,
url=https://ir.web.th-koeln.de
]
\address[1]{TH Köln (University of Applied Sciences),
  Gustav-Heinemann-Ufer 54, 50968 Köln, Germany}

\author[1]{Svetlana Myshkina}[%
%orcid=0000-0001-7116-9338,
email=svetlana.myshkina@smail.th-koeln.de,
%url=https://kmitd.github.io/ilaria/,
]

\author[1]{Jüri Keller}[%
orcid=0000-0002-9392-8646,
email=jueri.keller@th-koeln.de,
%url=https://kmitd.github.io/ilaria/,
]
\begin{abstract}
  The German Information Retrieval community is located in two different sub-fields: Information and computer science. There are no current studies that investigate these communities on a scientometric level. Available studies only focus on the information scientific part of the community. We generated a data set of 401 recent IR-related publications extracted from six core IR conferences from a mainly computer scientific background. We analyze this data set at the institutional and researcher level. The data set is publicly released, and we also demonstrate a mapping use case. 
\end{abstract}

\begin{keywords}
  Information retrieval \sep
  scientometric analysis \sep
  data set \sep
  co-authorship \sep
  research institutes \sep 
  networks
\end{keywords}

\maketitle

\section{Introduction}

Information Retrieval (IR) is an academic field with a long-standing history and a rather small but quite productive community. Traditionally, the German IR community is affiliated with information science and computer science, as it is an interdisciplinary field of research. The largest scientific organization with currently more than 200 members is the Special Interest Group for Information Retrieval (Fachgruppe Information Retrieval, FG IR)\footnote{\url{https://fg-retrieval.gi.de/}} of the German Computer Society (Gesellschaft für Informatik, GI), but also other related organizations like the Hochschulverband Informationswissenschaft (HI)\footnote{\url{https://www.informationswissenschaft.org}} with more than 100 members, have a notable amount of IR-related members. 

Scientific communities not only form and organize themselves with the help of scientific associations or societies but primarily by organizing conferences to bring together researchers, exchange ideas, and foster future developments in the field. The results of such conferences are published in conference proceedings. Therefore, to learn more about a scientific field and its participants, one can analyze the publication output of a community with the help of scientometric analyses, as shown in the IR-related bibliometric study of \citet{DING2001817} from 2000. More recent studies are scarce and mostly look at the information science side of the community or do not look at the German community specifically.

Therefore in this work, we will investigate some recent characteristics of the German IR community by building a data set of recent publications from six core IR conferences. We will analyze this data set at the institutional and the individual researcher levels to learn more about central actors in the field and to show first and preliminary results of this study.  

\section{Related Work}

An example of a German-focused study is the work of \citet{DBLP:conf/isiwi/BaumgartnerS07}, who analyzed the proceedings of the International Symposium on Information Science (ISI) between 1990 and 2004. They discussed the international claim of the conference with respect to academic coverage and the quality of the publications but did not focus on IR specifically. The analysis showed that the articles were written by 1.6 authors on average. This corresponds to the usual publication behavior in the information sciences. Most of the articles, 81\%, are written in German, but 57 of all cited references were in English. The evaluation of the ISI conference proceedings revealed a highly skewed distribution of authorship. The share of authors with more than three publications was only 4.8\%, while specific authors are constantly represented and form a core community. The most productive research groups came from Konstanz, Graz, Regensburg, Hildesheim, and Saarbrücken. 

In 2015, \citet{DBLP:conf/isiwi/LewandowskiH15} analyzed the citation behavior of German information scientists and studied the cited literature within the handbook ``Grundlagen der praktischen Information und Dokumentation'', which has a specific set of chapters related to IR topics. They found similar patterns to Baumgartner and Schlögl, by having an average of 1.4 authors per paper and mostly citing journal articles and writing in German. Considering the findings of \citet{https://doi.org/10.1002/asi.22645} that due to the specialization of authors in specific research areas, these can be representative of topics (concepts), they evaluate the results of their cluster analysis. They deduce that German-speaking information scientists probably work on somewhat distant topics. Consequently, the German-language information science community does not seem cohesive.

More tailored explicitly to the IR community, the bibliometric studies of \citet{DING2001817}, \citet{thornley_bibliometric_2011}, and more recently \citet{ferro_scholarly_2019} have to be mentioned. \citeauthor{DING2001817} used a co-word analysis to map the academic community in the field of information retrieval. While this work is highly recognized in the field with more than 1000 citations, it is not giving any insights into current trends, groups, or individuals, as it is based on a publication analysis from publications from before the year 2000. The analyses of \citeauthor{thornley_bibliometric_2011} and \citeauthor{ferro_scholarly_2019} are more recent but do not focus on Germany and also look at single conferences (TRECVid and CLEF) only. 

While conducting scientometric studies one has to keep in mind that these kinds of studies are often limited. Selection bias and inclusion/exclusion criteria are usual suspects when it comes to the validity of these kinds of studies \cite{archambault_benchmarking_2006}. These are not only introduced on the researcher level, who chooses which publication to include in the analysis but also on the data provider level \cite{DBLP:conf/birws/MichelsNSS20}. Database curators can influence the coverage of the databases scientometric researchers work on and therefore influence the results.

\section{Data Set Generation}

We compiled a data set of research publications by analyzing the proceedings of six major IR-related and peer-reviewed conferences (CHIIR, CIKM, CLEF, ECIR, SIGIR, WWW) available in the ACM Digital Library. The publication dates ranged between January 2020 and June 2023. We only considered those publications that had at least one German author or an author that was affiliated with a German research institute. Other central publication venues of the community, like the Springer Information Retrieval Journal or ACM TOIS were not considered, as we wanted to focus on conferences. The TREC and CLEF workshop notes were also not included due to the missing peer review. 

For the six conferences mentioned above, we gathered the following publications' metadata:
\begin{itemize}
    \item author names,
    \item affiliation,
    \item titles,
    \item DOI of the publication.
\end{itemize}

As affiliation names were given in many different variants, we harmonized them manually. We added detailed information about the working group as well as the complete postal address and geo-location.
In the case of different departments of research groups within an institute, we kept all sub-groups separate as long as the sub-group names were explicitly mentioned in the publications.
In total, we discovered 401 publications and 195 different affiliations. 
The data set is publicly available in a GitHut repository\footnote{\url{https://github.com/irgroup/LWDA2023-IR-community}}.

%Bei der Aufbereitung der erhobenen Daten werden die zurückgelieferten Treffermengen gesichtet und in Auswahl manuell in eine Auswertungstabelle (Excel) überführt. Dem Auswahlverfahren liegen zum einen die oben genannten Erhebungskriterien zugrunde. Zum anderen ist pro Publikation jede einzelne Institution nur einmal zählerisch zu erfassen. (Ausnahmen gelten bei Vorkommen von Zugehörigkeit zu mehreren Institutionen. In diesen Fällen soll gesondert verfahren werden.)
%Zu den pro Veröffentlichung erhobenen Elementen gehören: Nach- und Vorname des Erstautors (bzw. der Autoren), Institution, DOI der Publikation. Einzelne Arbeitsgruppen oder Institute der jeweiligen Einrichtungen sowie ihre Adressen sind durch punktuelle Nachrecherchen zu ermitteln.

\section{Most Productive Research Groups }

From the 195 research groups we found in the data set, we extracted the ten most productive ones (see Table \ref{tab:research_groups}). The most productive group is the Webis Group from Weimar, Leipizg, Jena, and formerly Halle, followed closely by the Databases and Information Systems group from Max Planck Institute and the L3S research center. Webis and L3S are ``virtual groups'' whose members also have co-affiliations with universities or non-university research centers (like TIB Hannover). We decided to not split up these publications, as the publishing collaboration within these institutes is intense, and authors publicly affiliate with them. Due to the size of these groups, it's not surprising to see them at the top of the list. If we would have split up these groups, the ordering would not have changed a lot, as the two central locations of Webis Weimar and Leipzig would still have made it to the top. 

The list also includes one commercial research institute (Bosch Center for Artificial Intelligence), one university of applied sciences (TH Köln), and two non-university research centers (GESIS and TIB), showing the interdisciplinary and heterogeneous constitution of the German IR community. 

Next to the sum of all publications within the time frame, we separated the count for each of the six core conferences. We also see a mixed set of publication profiles. The University of Regensburg only published at CHIIR and ECIR, while Max Planck, Webis, and L3S had a more heterogeneous coverage. 

\begin{table}[t]
    \centering
    \caption{Most productive German research groups and their publication counts between 2020 and 2023.}
    \begin{tabular}{lrrrrrrr}
    \toprule
    {} & Total &  \rotatebox{90}{SIGIR} &  \rotatebox{90}{CIKM} &  \rotatebox{90}{WWW} &  \rotatebox{90}{ECIR} &  \rotatebox{90}{CHIIR} &  \rotatebox{90}{CLEF} \\
    \midrule
    Webis Group                                        &                        37 &      6 &     6 &    0 &    13 &      5 &     7 \\
    Max Planck Institute - Databases and Inf. Systems  &                        30 &     13 &     7 &    4 &     5 &      1 &     0 \\
    Forschungszentrum L3S                              &                       28 &      3 &    10 &   14 &     0 &      1 &     0 \\
    GESIS - Leibniz-Institut für Sozialwissenschaften  &                     13 &      1 &     4 &    3 &     0 &      5 &     0 \\
    TIB - Forschungsgruppe Visual Analytics            &                      13 &      3 &     3 &    3 &     2 &      2 &     0 \\
    U Bonn - Data Science \& Intelligent Systems       &                       11 &      1 &     6 &    3 &     0 &      1 &     0 \\
    U Regensburg - Chair of Information Science        &                        10 &      0 &     0 &    0 &     3 &      7 &     0 \\
    U Mannheim - Data and Web Science Group            &                       10 &      0 &     4 &    5 &     0 &      1 &     0 \\
    Bosch Center for Artificial Intelligence           &                       10 &      1 &     5 &    4 &     0 &      0 &     0 \\
    TH Köln - Information Retrieval Research Group     &                        9 &      2 &     0 &    0 &     3 &      0 &     4 \\
    \bottomrule
    \end{tabular}
    \label{tab:research_groups}
\end{table}

\begin{table}[t]
    \centering
    \caption{Composition of co-authorships for the ten most productive IR groups in Germany 2020-2023.}
    \begin{tabular}{lrrr}
    \toprule
    {} &  Authors$_{min}$ &  Authors$_{mean}$ &  Authors$_{max}$ \\
    \midrule
    Webis Group                                        &            3 &             8 &           17 \\
    Max Planck Institute - Databases and Inf. Systems &            1 &             3 &            6 \\
    Forschungszentrum L3S                              &            1 &             4 &           12 \\
    GESIS - Leibniz-Institut für Sozialwissenschaften  &            3 &             5 &           12 \\
    TIB - Forschungsgruppe Visual Analytics            &            3 &             5 &           12 \\
    U Bonn - Data Science \& Intelligent Systems       &            2 &             5 &           12 \\
    U Regensburg - Chair of Information Science          &            1 &             3 &            6 \\
    U Mannheim - Data and Web Science Group              &            2 &             3 &            6 \\
    Bosch Center for Artificial Intelligence           &            3 &             6 &           10 \\
    TH Köln - Information Retrieval Research Group     &            2 &             4 &            7 \\
    \bottomrule
    \end{tabular}
    \label{tab:research_groups-authors}
\end{table}

\begin{table}[t]
    \centering
 \caption{Thirty authors with the highest betweenness centrality in the undirected co-authorship network and their number of publications between 2020-01 and 2023-06.}    
\begin{tabular}{llll}
\toprule
Author               & Affiliation & Publications & Betweenness\\
\midrule

Lucie Flek                & U Bonn / U Marburg                                                   & 3            & 0.007333        \\
Martin Potthast           & Webis Group                                                           & 27           & 0.005312        \\
Ralph Ewerth              & TIB Hannover                                                          & 7            & 0.005023        \\
Benno Stein               & Webis Group                                                           & 26           & 0.004141        \\
Jens Lehmann              & Amazon                                                                & 7            & 0.003524        \\
Stefan Dietze             & GESIS, Köln                                                           & 7            & 0.003436        \\
Gerhard Weikum            & Max Planck Institute                                    & 12           & 0.003282        \\
Avishek Anand             & L3S                                                                   & 8            & 0.002947        \\
Rishiraj Saha Roy         & Max Planck Institute                                    & 7            & 0.002798        \\
Daniel Hienert            & GESIS, Köln                                                           & 4            & 0.002675        \\
Kuldeep Singh             & Cerence                                                               & 6            & 0.002645        \\
Megha Khosla              & L3S                                                                   & 3            & 0.002468        \\
Norbert Fuhr              & U Duisburg-Essen                                            & 7            & 0.002403        \\
Andrew Yates              & Max Planck Institute                                    & 12           & 0.002078        \\
Axel-Cyrille Ngonga Ngomo & U Paderborn                                                 & 3            & 0.001734        \\
Matthias Hagen            & Webis Group                                                           & 21           & 0.001447        \\
Ran Yu                    & GESIS, Köln                                                           & 7            & 0.001362        \\
Chris Biemann             & U Hamburg                                                   & 7            & 0.001316        \\
Sherzod Hakimov           & U Potsdam                                                   & 4            & 0.001148        \\
Maria-Esther Vidal        & L3S                                                                   & 3            & 0.001042        \\
Endri Kacupaj             & Cerence                                                               & 3            & 0.000786        \\
Henning Wachsmuth         & Webis Group                                                           & 8            & 0.000759        \\
David Elsweiler           & U Regensburg                                                & 6            & 0.000694        \\
Andreas Both              & HTWK Leipzig                                                          & 2            & 0.000667        \\
Janek Bevendorff          & Webis Group                                                           & 9            & 0.000667        \\
Maria Maleshkova          & U der Bundeswehr, Hamburg & 2            & 0.000623        \\
Philipp Schaer            & TH Köln                                                               & 8            & 0.000601        \\
Timo Breuer               & TH Köln                                                               & 6            & 0.000601        \\
Alexander Bondarenko      & Webis Group                                                           & 12           & 0.000600        \\
Yvonne Kammerer           & HDM Stuttgart                                                         & 5            & 0.000576 \\
\bottomrule
\end{tabular}

    \label{tab:co-authorship}
\end{table}

\section{Co-Authorship in the German IR Community}

In contrast to the previously mentioned observations from the field of information science, we wanted to check on the characteristics of co-authorship and collaboration in more computer science-related IR conferences. In Table \ref{tab:research_groups-authors} the number of authors ranges from 1 to 17 for single publications. On average, the top ten research groups published papers with 4.83 authors, while on the whole data set the average number of authors was 4.98. With eight authors the Webis Group has the highest number of authors per paper on average, while the second most productive group (Max Planck) has the lowest number of authors per paper (three on average). The number of authors alone can therefore not explain the publication success of a group. We should nevertheless keep in mind that the number of co-authors per paper can introduce distortions for the calculation of additional network-based performance metrics \cite{masic_inflated_2021}.

Given these limitations, we analyzed the publications with the help of a co-authorship network. All author collaborations form a network of 1159 nodes (authors) and 4907 edges (co-authorship relations). To allow a more nuanced impression in comparison to simple publication counts, we calculated betweenness centrality on the network. Betweenness centrality tells us how important a node is to form the network by connecting different parts. In a co-authorship network, these are authors that bridge different communities of working groups and are important for the connectivity of the network. 

In Table \ref{tab:co-authorship}, we see the top thirty best-connected authors of the field. We only selected those researchers who work at German research institutes. However, we can see in the co-authorship network that many other well-connected researchers are from outside of Germany. The top-ranked researcher is Lucie Flek (U Bonn and U Marburg), with only three publications in total, which might be surprising, but these papers were published at WWW, SIGIR, and ECIR and had no single overlap in co-authors. Therefore she is a well-connected author in the co-authorship network with only three papers. Other well-connected authors, like Martin Potthast and Benno Stein (both Webis Group) had a much larger number of publications, in comparison. 

%\begin{figure}
%    \centering
%    \includegraphics[width=\linewidth]{co-author-network.png}
%    \caption{Co-authorship network of 1159 authors from 401 IR-related publications.}
%    \label{fig:enter-label}
%\end{figure}

\section{Topics of Publications per Research Group}
We applied a simple topic modeling of the top ten research institutes based on the titles of the publications published by each institute. The titles are combined into documents and the terms in these documents up to bi-grams are weighted by TF-IDF after a basic text processing. The top terms per institute with the highest TF-IDF give insights about the topics the institute mainly focuses on, and which differentiates it from the other groups. Table~\ref{tab:topics-top5} gives an overview of the top 3 terms per research institute. While most terms like \textit{knowledge}, \textit{question answering} or \textit{search} are related to IR, some terms appear unexpected. The highest ranked term for the Bosch Center for Artificial Intelligence is, for example, \textit{welding}. Since this term is rarely used in the context of IR, in two publications the institute describes how welding can be monitored through the help of IR techniques~\cite{DBLP:conf/cikm/ZhengZZSK22a, DBLP:conf/cikm/ZhouSBPMK20}.

Likewise, the title terms of the publications with German authors were analyzed by conference. By comparing the top terms of the groups with the top terms of the conference, a group mainly publishes some correlations could be found. For example, for the Max Planck Institute with the Databases and Information Systems group \textit{question answering} has a high TF-IDF, which ranks high for the SIGIR conference in which the group mainly publishes. Other indirect similarities can be observed for example between the L3S with machine learning-related terms and the CIKM. Since the initial data set for the topics of the conferences and groups is the same and, therefore productive groups significantly influence the top terms per conference, a correlation is not surprising. 

\begin{table}
    
    \caption{Top 3 terms from the titles of the publications of the research institutes ranked by TF-IDF.}
    \label{tab:topics-top5}
    
\begin{tabular}{llll}
\toprule
{} &                 1 &                        2 &                   3 \\
\midrule
Webis Group                                       &          overview &                 argument &              touché \\
Max Planck Institute &         answering &  conversational question &  question answering \\
Forschungszentrum L3S                             &            neural &                    using &             forward \\
GESIS  &  language queries &                knowledge &      knowledge base \\
TIB            &        multimodal &              geolocation &              search \\
U Bonn       &         knowledge &          knowledge graph &               graph \\
U Regensburg       &          snippets &        featured snippets &            featured \\
U Mannheim           &          matching &                detection &               using \\
Bosch           &           welding &                  machine &    machine learning \\
TH Köln     &       experiments &           ir experiments &                  ir \\
\bottomrule
\end{tabular}
\end{table}

\section{Visualizing IR research groups}

\begin{figure}[t]
    \centering
    \includegraphics[width=\linewidth]{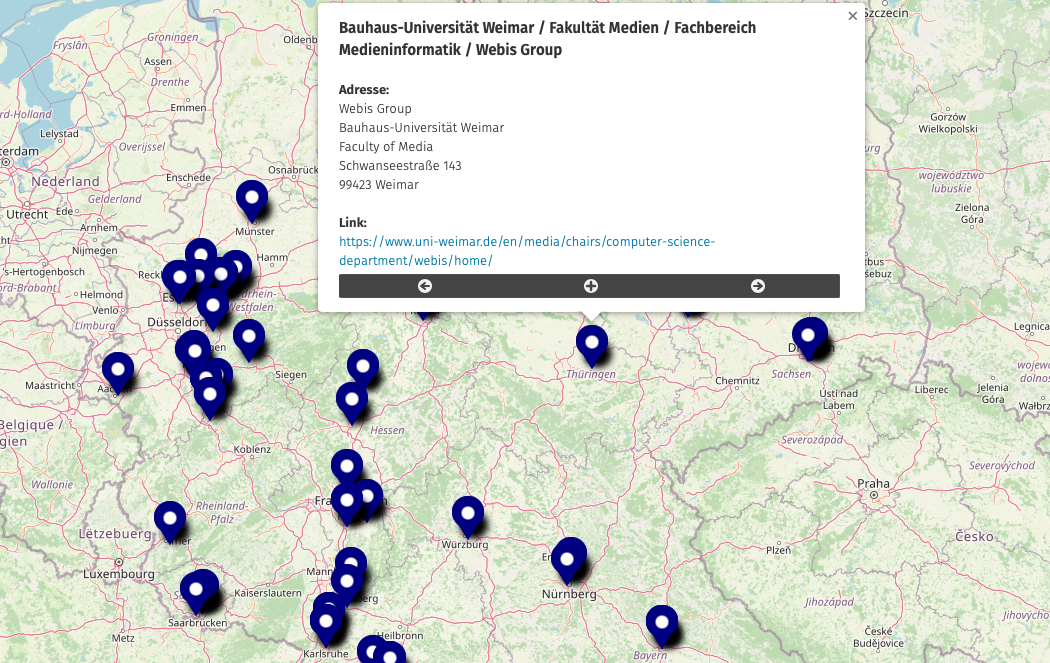}
    \caption{Dynamic map visualization of all German research institutes that were published between 2020 and 2023 in six major IR venues (CHIIR, CIKM, CLEF, ECIR, SIGIR, WWW).}
    \label{fig:ir-map}
\end{figure}

We use the available data to draw the geo-locations of each research group on a map using the OpenStreetMap platform. The map is available online\footnote{\url{http://u.osmfr.org/m/931946/}}. The map includes all groups and institutes that published at least two times in the six previously mentioned conferences. To extend the map with additional institutes that might be missed, we added groups that were active in academic societies, like the German Special Interest Group on Information Retrieval between 2020 and 2023. Each data point on the map includes the name of the group, the postal address, and a link to the institute's homepage. Figure \ref{fig:ir-map} shows a sample of the map.

\section{Discussion and Future Work}

We conducted a small-scale scientometric study of the German IR community using publications from 2020 till mid-2023. For a more in-depth investigation, we would need a more extended time coverage and should investigate more conferences to reflect the field's heterogeneity. Additionally, the chosen time frame mainly consists of the COVID-19 pandemic, which might have introduced some uncommon publication patterns (like submitting to conferences without traveling to these conferences). The six conferences are relevant to the field, but other related conferences like JCDL or the ICTIR might also be included in a later version of the data set. 
Including CIKM might have introduced some topical shift in the data set, as the main focus of CIKM is not information retrieval (although some relevant IR papers are located there). A more fine-tuned topical selection process might increase the quality of the data set. Additionally, the decision to discard TREC and CLEF lab notebooks also excluded some potentially interesting publications and, consequently, may be the reason for missing research groups.  

While all these limitations are known and are valid complaints regarding the preliminary results of this scientometric study, it's the only available data set for the German IR community. The dataset in its current form therefore only allows us to get some first and preliminary insights into the community, and its actors on the institutional and researcher levels. Nevertheless, it gives us an idea of the rich collaborations happening in the field. 

\bibliography{references}
\end{document}